\documentclass{JAIS}

\journal{JAIS-ID}
\vol{2022}

\received{xx January 2018}
\published{xx March 2018}

\def\be{\begin{equation}}
\def\ee{\end{equation}}
\def\bea{\begin{eqnarray}}
\def\eea{\end{eqnarray}}

\usepackage{caption}
\usepackage{lineno}

\usepackage{float}
\usepackage{graphicx} % Required for including images
\usepackage{subcaption} % Required for subfigures
\usepackage{hyperref}

\articletype{}

\begin{document}

\title{Mapping Water on the Moon and Mars using a Muon Tomograph}

\author{Olin Lyod Pinto\auno{1,2}, Jörg Miikael Tiit\auno{2,3}}
\address{$^1$National Institute of Chemical Physics and Biophysics, Tallinn, 12618, Estonia \\
$^2$GSCAN ÖU, Maealuse2/1, Tallinn, 12618, Estonia \\ 
$^3$ STACC ÖU, Narva mnt 20, Tartu, 51009, Estonia}
\address{
\vspace{0.2cm}
Corresponding author: Olin Lyod Pinto\\
Email address: olin.pinto@kbfi.ee
}

\begin{abstract}
The search for water on the Lunar and Martian surfaces is a fundamental aspect of space exploration, contributing to the understanding of the history and evolution of these celestial bodies. However, the current understanding of the distribution, concentration, origin, and migration of water on these surfaces is limited. Moreover, there is a need for more detailed data on these aspects of Lunar and Martian water. The natural flux of cosmic-ray muons, capable of penetrating the planetary surface, offers a method to study the water-ice content, composition, and density of these surfaces. In this paper, the author presents a novel approach to address these knowledge gaps by employing cosmic-ray muon detectors and backscattered radiation. The study describes a cutting-edge muon tracking system developed by GScan and highlights the results of preliminary simulations conducted using GEANT4. These findings suggest that muon tomography could be a potential tool for investigating water-ice content on the Lunar and Martian surfaces, pointing to new avenues for space science exploration.
\end{abstract}

\maketitle


\section{Muography in space}
The exploration of celestial bodies beyond Earth has captivated the curiosity of scientists and space enthusiasts for centuries. One fundamental aspect of such exploration is the search for water, a vital resource for sustaining life and enabling future human missions. In recent years, the Moon and Mars have emerged as primary targets for this quest, with extensive efforts to understand the distribution and nature of water on these celestial bodies.

Historically, the Moon was regarded as a desiccated and arid world without significant water resources. However, a paradigm shift occurred in 2007 when scientific observations hinted at the presence of water in the Lunar mantle. The Lunar Prospector mission, launched in 1998, detected elevated hydrogen concentrations at the Moon's poles, suggesting the existence of water~\cite{binder1998lunar}. Subsequent missions, such as Chandrayaan-1 in 2009, confirmed the presence of surface water-ice in certain permanently shadowed craters near the Lunar poles~\cite{priyadarshini2009water}. The LCROSS impact experiment further estimated the water content in the Moon's regolith to be 5.7\% by weight~\cite{colaprete2010detection}. More recently, the Stratospheric Observatory for Infrared Astronomy (SOFIA) detected molecular water in the illuminated regions of the Moon~\cite{honniball2021molecular}. Despite these exciting findings, the mechanisms responsible for water containment within the Lunar and Martian subsurface and the potential extraction methods remain subjects of ongoing debate and investigation.

Like the Moon, Mars has also been a target for water exploration. Various forms of water, including ice in the polar caps, glaciers at lower latitudes, and subsurface permafrost, have been detected on the Martian surface~\cite{nazari2020water}. These discoveries have sparked interest in understanding the extent and distribution of Martian water resources, as they hold implications for future human colonization and sustained exploration on the Red Planet.

To address the challenges of finding water on the Moon and Mars, novel techniques and instrumentation are being developed. One such method gaining prominence is muon tomography, a non-invasive imaging technique that utilizes cosmic rays to probe the interior structure of planetary bodies. Muons, which are high-energy particles generated by cosmic ray interactions on the Lunar or Martian surface, can penetrate significant depths and may backscatter. By measuring the attenuation and backscattering of muons, we can infer the presence and distribution of water within these celestial bodies.

The motivation for utilizing muon tomography in space exploration lies in its ability to provide valuable insights into the subsurface composition of the Moon and Mars. This non-destructive technique offers the potential to precisely map the distribution of water and other dense materials, allowing us to better understand the geological processes and history of these planetary bodies. Moreover, the knowledge gained from muon tomography studies can aid in identifying regions with higher water concentrations, facilitating the planning of future missions and resource utilization.

From a scientific standpoint, deploying a muon tomography instrument serves many possibilities for various investigations and avenues of research. As an example, drilling into the sub-surface poses insurmountable challenges. However, the problems of drilling could likely be overcome relatively easily by using the natural flux of the cosmic muons (or other charged particles, as well as neutrons). This approach would allow for a faster, more efficient, and possibly higher accuracy analysis compared to traditional physical sample methods. Furthermore, due to the high penetration capabilities of cosmic ray particles, the muon tomography methods carry great potential for gathering information from deeper layers than drills can reach.

This article aims to present a study using the muon tomography method in space. It discusses the current understanding of Lunar and Martian water resources, the significance of utilizing muon tomography in space exploration, and the potential implications of the findings. This research contributes to a broad understanding of the feasibility of finding water content on planetary surfaces and for future exploration endeavours.

\section{Detector Technology}

\begin{figure}[H]
\centering
\begin{subfigure}{0.45\textwidth}
\includegraphics[height=6.cm, width=0.8\textwidth]{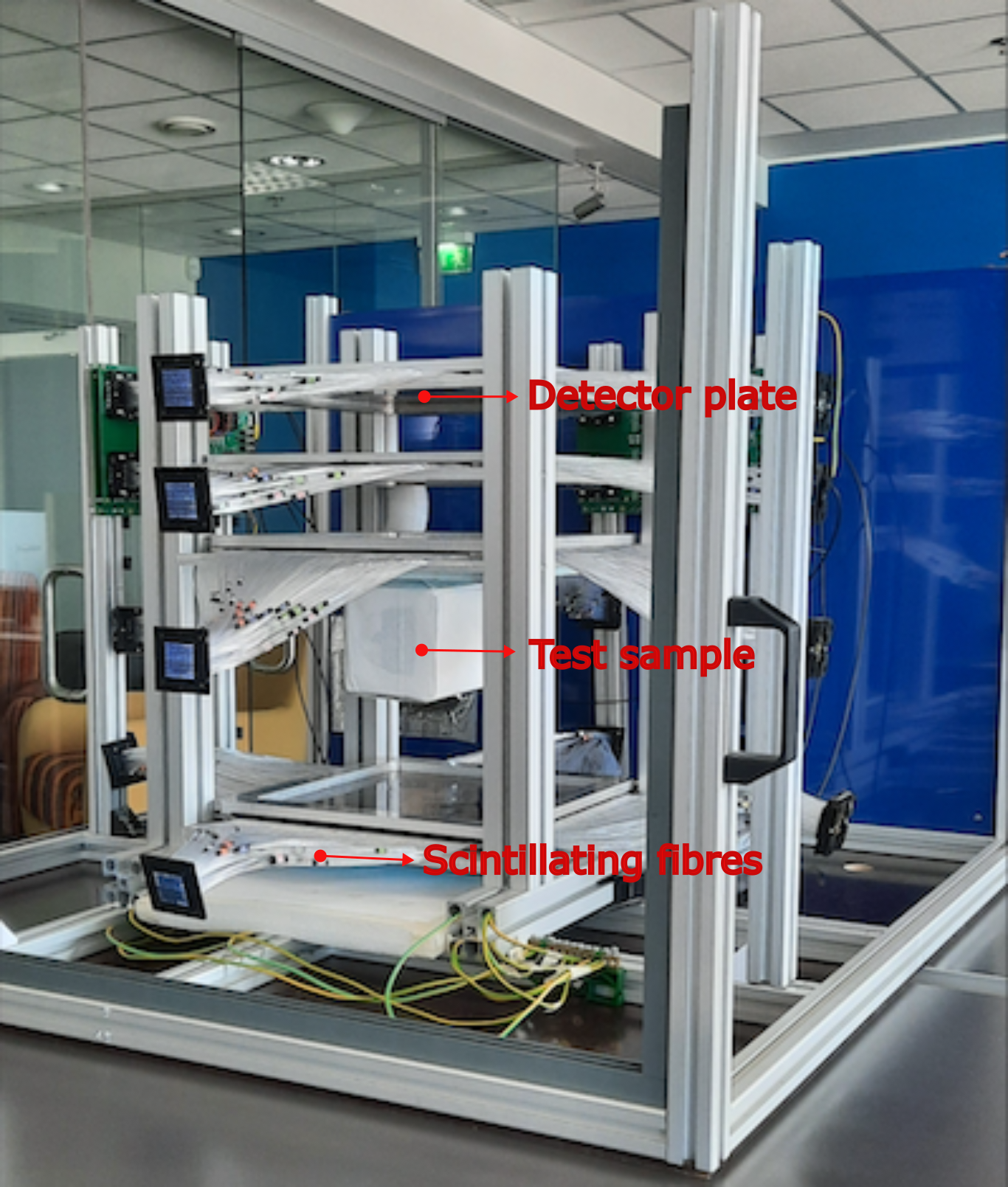}
\caption{}
\label{subfig:labelledMVP}
\end{subfigure}
\begin{subfigure}{0.45\textwidth}
\includegraphics[height=6.cm, width=\textwidth]{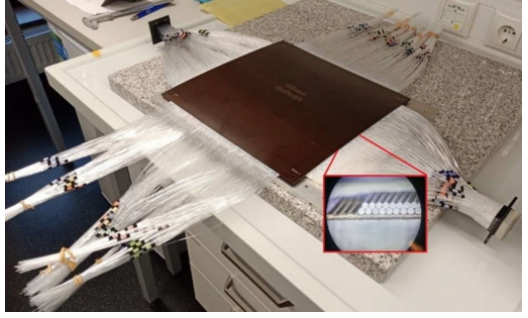}
\caption{}
\label{subfig:Layer}
\end{subfigure}
\caption{a) A photo of the GScan prototype muon tomography consisting of four detector plates (left) in a 3-1 configuration with a volume of interest at the center and b) The assembly of a detector plate. Figure is taken from~\cite{gscan1}}
\label{fig:GSCAN_MVP}
\end{figure}

The GScan Muon Tracker is based on scintillating fibres, as they are widely used and offer several advantages, making them suitable for a wide variety of applications. The GScan Muon Tracker is a 4-layered prototype hodoscope (3-1 configuration) with 1 mm Saint-Gobain BCF-12 single cladding scintillation fibres~\cite{gscan1}. The fibres are arranged as two double-layered fibre mats oriented orthogonally to each other, as shown in~\ref{fig:GSCAN_MVP}. This configuration ensures the angular resolution of the system of about one milliradian. The three hodoscope layers in the top part are separated by 75 mm, with the lowest detector plate set 250 mm from the last of the top layers. The configuration provided a 247 mm $\times$ 247 mm active area for every hodoscope, hence a total volume of interest (VOI) of 247 mm $\times$ 247 mm $\times$ 250 mm$^3$. The data acquisition system consisted of eight CAEN DT5550W boards paired with Ketek (PA3325-WB-0808) and Hamamatsu (S13361-3050AE-08) SiPMs arrays for collecting the scintillation light from the fibres.

It is noteworthy to mention that a comparable strategy is being pursued in the domain of space exploration, specifically with regard to muon tomography. In this context, the approach is distinct, as it exclusively encompasses the utilization of the topmost three detector layers. This refinement is calibrated to align with the unique demands of space-based applications. The table~\ref{table:lunar_detector_specs} outlines the critical specifications and general requirements envisioned for a muon tomograph tailored for space missions.

\begin{table}[H]
\centering
\begin{tabular}{ll}
\hline
\textbf{Parameter} & \textbf{Details} \\
\hline
Detector Material & Polystyrene-core scintillating fibres (0.5mm - 1mm diameter) \\
Detector Geometry & total size $\sim$1m$^3$; 10 cm gap between layers,\\
 & alternating fiber orientation; minimum of 3 layers \\
Resolution & Position (1mm);  Angular ($\sim$1 degree); Time (50 -100 ps)\\
Readout Electronics & Radiation-hardened Silicon Photomultipliers (450-550 nm) \\
Support Frame & Aluminium or Titanium \\
Data Processing & Field Programmable Gate Arrays (FPGAs) \\
Power System & Solar panels with Radioisotope Thermoelectric Generators (RTGs) \\
Shielding & Polyethylene for electronics and SiPMs \\
\hline
\end{tabular}
\caption{Specifications and requirements for a muon tomograph designed for a space mission.}
\label{table:lunar_detector_specs}
\end{table}

\section{Methodology}

The core methodology of this study is hinged on evaluating the potential of utilising upward-travelling muons, in coordination with a strategically-positioned GScan tracker on the planetary surface, to quantify water content in the soil. At the heart of this methodology lies the unique behaviour of muons, which are generated through interactions between cosmic rays and the planetary surface. These highly energetic particles possess the ability to penetrate deep into geological formations, thus serving as instrumental probes for the identification and subsequent in-depth characterisation of underlying structures. Concurrently, the GScan tracker, specifically engineered for muon tracking, functions as an essential apparatus for this study.

The GScan muon tracker is reconfigured, incorporating a top hodoscope. This methodology also emphasises the critical importance of grazing muons, which, due to their shallow interaction angles with subsurface formations, act as sources of backscattered particles, causing scattering events and directional alterations. In order to compile statistically significant data regarding upward muon flux, integrating a large-scale detector is essential. Optimally reflecting the dimensions of the Lunar or Martian terrain under examination, this expansive detector ensures the collection of a considerable number of upward-travelling muons and their interactions with subsurface geological entities. This methodological approach minimises statistical uncertainties, enhancing the overall quality of precision, accuracy, and dependability of the outcomes derived from muon tracking experiments.

The focus of this study extends to specific variables such as hit energy, time, and scattering angle distributions. Hit energy is defined as the energy deposited by the particle in the sensitive volume of the detector; in this case, it is a scintillating plate. Hit time is the time recorded between the detector planes are t$_3$ - t$_2$ $>$ 0 and t$_2$ - t$_1$ $>$ 0. The scattering angle is obtained from the scattering angle $\theta$ between a reference direction vector (A) and a momentum direction vector (B) is:

\begin{equation}
 \theta = \frac{arccos(A \cdot B)}{|A|*|B|}   
\end{equation}

A thorough examination is conducted through a comparative investigation of these distributions across Lunar and Martian surfaces and further encompassing conditions involving dry soil and underlying frozen lake (hereby referred to as ``Rock'' and ``Ice'').

\section{Simulations}

The preliminary simulation aimed to establish a fundamental understanding of the potential shortcomings and the overall feasibility of the initially chosen simulation concept. The simulation was performed using the GEANT4 framework~\cite{agostinelli2003geant4} to simulate a simplified Lunar/Martian environment and the passage of particles through both celestial bodies. GEANT4's QGSP\_BERT physics list was used to model the interactions of particles with matter.

\smallbreak

\begin{figure}[H]
    \centering
    \begin{subfigure}{0.4\textwidth}
        \includegraphics[height=6.5cm, width=\textwidth]{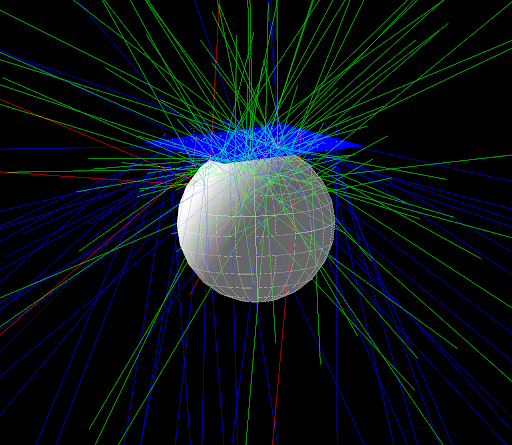}
        \caption{}
        \label{subfig:Lunar_Sim}
    \end{subfigure}
    \begin{subfigure}{0.4\textwidth}
        \includegraphics[height=6.5cm, width=\textwidth]{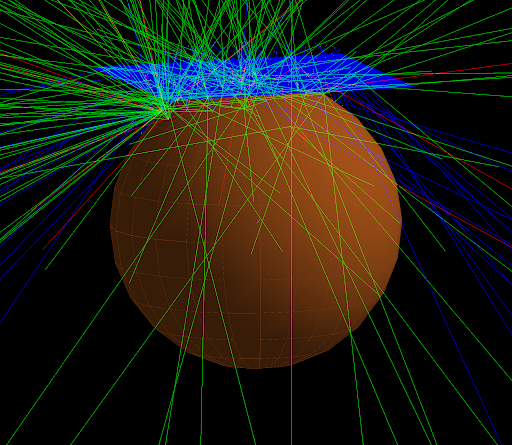}
        \caption{}
        \label{subfig:Mars_Sim}
    \end{subfigure}
    \caption{Simulations a) the Moon and b) Mars. Cosmic rays are coming from the top view. Protons or alpha particles are shown in blue, while secondary particles are represented in green. The Martian scenario does not account for magnetic fields or atmospheric effects.}
    \label{Fig:SimModel}
\end{figure}

For the simulation, three scintillator plates (top hodoscope), each with a thickness of 0.2 cm, were used as the active material for particle detection, with the plates placed 10 cm apart to capture the trajectories effectively - the distance and thickness were chosen to replicate the GScan's muon tomography setup. The use of a large-scale detector, matching the diameter of the Moon/Mars, was essential to obtain statistically significant data, improving accuracy and reliability. This large-scale detector enables capturing a substantial number of upward-travelling particles and their interactions with subsurface geological formations. Moreover, it is considered an effective area; therefore, it reduces the amount of computational time significantly. However, the implications of this choice, such as how much exposure time would actually be needed with a realistically sized detector as mentioned in table~\ref{table:lunar_detector_specs} to achieve enough statistics, must be considered and warrant further investigation. The visualisation of the Moon and Mars model and the detector can be seen in Figure~\ref{Fig:SimModel}.

As the cosmic ray source (Galactic Cosmic Rays, or GCR), the EcoMug model~\cite{pagano2021ecomug} was employed, which approximates the primary cosmic ray flux based on the defined parameters. The composition of the cosmic ray flux was assumed to consist of 85\% protons and the remaining 15\% of alpha particles - reflecting the average composition observed in cosmic ray measurements. The zenith angle (angle of approach to the Lunar/Martian surface) was limited to 75 degrees to obtain an increased amount of grazing particles from the surface. The source height was set at 2000 km and 4000 km from the surface of the Moon and Mars. The energy range of the simulated particles spanned from 1 GeV to 100 TeV, covering a broad spectrum of commonly observed energies in cosmic ray measurements~\cite{ginzburg2013origin} - energies beyond 100 TeV are not simulated due to limitations from GEANT4 particle interactions.

\smallbreak

The Moon and Mars were modelled as simple geometrical spheres with radii of 1737.4 km and 3390 km, respectively. To accurately depict the surface variations of these two planetary environments, each spherical body was divided into three distinct layers: the crust, mantle, and core. The regolith (topmost surface) for the Moon and Mars was approximately 5 meters thick. Though the Martian regolith was also considered, its specific thickness was modelled according to observed data~\cite{anderson1979analysis, ming2017chemical}. The intention behind simulating these diverse layers was to analyse the interactions of particles with the various materials found in these celestial bodies and consequently, to explore the potential for detecting water in different regions. A realistic portrayal of muon production was achieved, with the majority occurring within 1 meter of both the Lunar and Martian soil.

\begin{figure}[H]
    \centering
    \begin{subfigure}{0.4\textwidth}
        \includegraphics[height=5.cm, width=\textwidth]{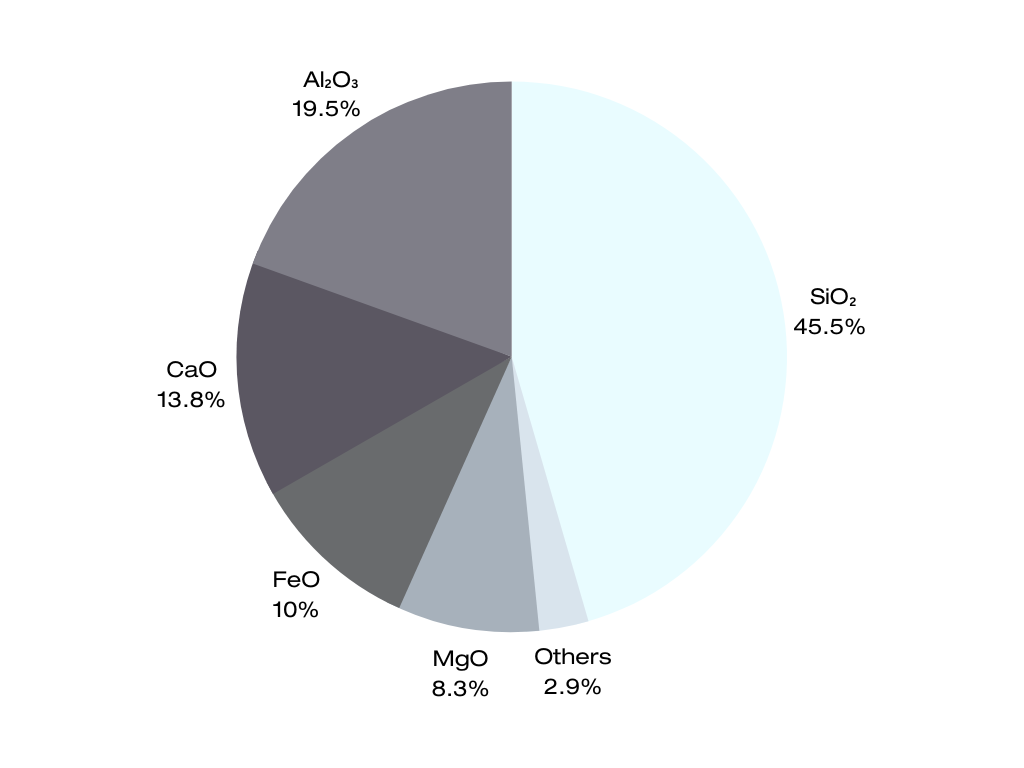}
    \end{subfigure}%
    \begin{subfigure}{0.4\textwidth}
        \includegraphics[height=5.cm, width=\textwidth]{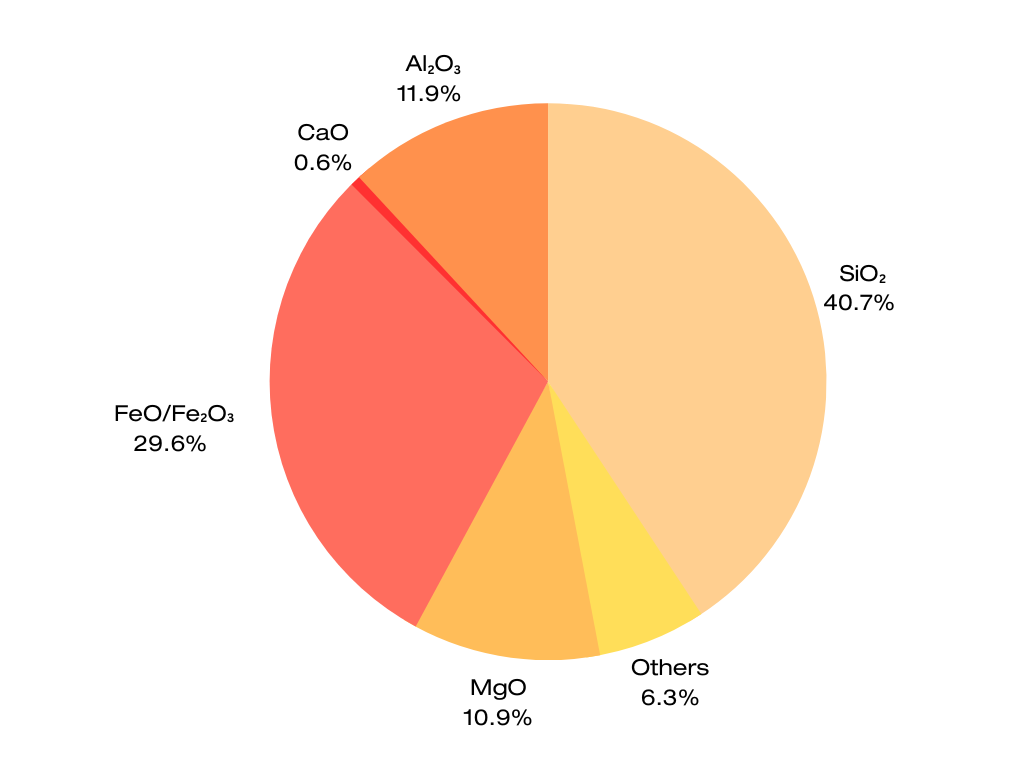}
    \end{subfigure}
    \caption{Chemical composition of Lunar and Martian surface~\cite{ming2017chemical, taylor2016lunar}.}
    \label{fig:chemcomp}
\end{figure}

The chemical composition for simulating the Lunar and Martian soil is shown in figure~\ref{fig:chemcomp}, and the simulations included two scenarios - the "dry" Lunar surface (Anorthite Rock) and a case in which a frozen lake 7 km deep at the top of the planetary surface. As the parameters of interest, 5-dimensional information (position in the $x$, $y$, $z$ directions, energy (E), and time (t)) and scattering angle are captured in addition to recording the particle types as the ground truth information. The collected information is initially analysed using the ROOT framework~\cite{brun1997root}, following which further analysis can be performed in Python. The primary goal is to differentiate the backscattered particles from those of forward-travelling and characterise the parametrical differences between the backscattered particles originating from the dry Lunar/Martian surface and those from the frozen lake scenario. Because the particles interact characteristically depending on the media of interest (tied to physical and chemical alterations of the media) and hence carry information about it, the patterns in the parameters of interest give rise to the detection capabilities - the parameter space distributions between the dry soil and frozen lake scenario, in this particular case, should be noticeably different from one another. 

Simulating the production of muons (and any other particles of interest produced within any given media) presents a significant challenge, making high-precision modelling of the Lunar/Martian surface imperative. Unlike the Earth, the Lunar environment lacks an atmosphere, causing primary particles to interact directly with the topmost layers of the surface. This direct interaction leads to the generation of muons (and other secondary particles), which differs from the process on Earth; where the majority of muons are produced after interactions with the atmosphere. Consequently, the backscattered muon flux on the Moon is notably smaller compared to Earth, as the absence of atmosphere-produced muons contributes to the reduced overall backscattered flux in the Lunar environment. This distinction in muon production mechanisms is crucial to be considered, helping to develop an improved understanding of the unique aspects of the Lunar surface and its particle interactions - properly accounting for these factors is essential for obtaining accurate results in muon simulations and subsequent analysis. 

\section{Data Analysis}

In the simulation of Galactic Cosmic Ray (GCR) events, a total of half a million interactions were investigated, focusing on the properties and distribution of backscattered particles. Out of this dataset, the detectors recorded an average of 10\% of the interactions as backscattered events. Photons, as the most prevalent component, constitute half of the backscattered particles, reflecting their significant role in the interaction process. Electrons follow, accounting for 10\% of the occurrences, illustrating their substantial presence among the backscattered particles. Protons, known for their stability and significant mass, were observed at a rate of 4\%. At the same time, pions were slightly more prevalent at 6\%, suggesting a nuanced relationship between these charged particles in the scattering phenomenon. Remarkably, muons were detected at a rate of less than 1\%, indicating their rare occurrence in the backscattering process despite being essential secondary products in cosmic ray interactions. Neutrons, with a recording rate of a mere 0.01\%, were the least prevalent among the backscattered particles.

\subsection{Spectral Analysis}
\begin{figure}[H]
\centering
\includegraphics[height=5.5cm, width=0.9\textwidth]{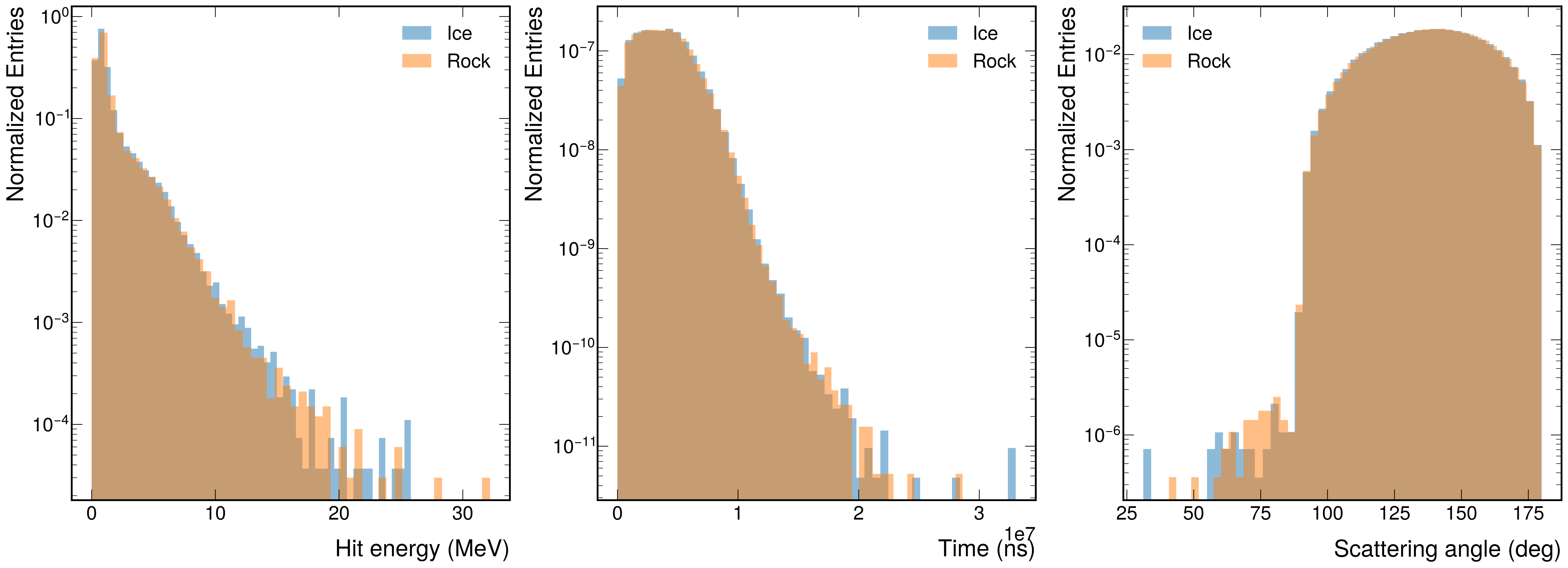}
\caption{Distribution of Hit energy, time and scattering angle, in comparison between Rock and Ice datasets in the context of the Lunar environment.}
\label{fig:RockIce_EnergyTimeAngle_All}
\end{figure}

Figure~\ref{fig:RockIce_EnergyTimeAngle_All} shows the hit energy, time and scattering angle distributions of the detected backscattered particles. These distributions are compared between two distinct materials: Ice and Rock.  Furthermore the distribution are extracted for each particle types. We see some discrepancy in shape between the datasets from Ice and Rock.

The comparison cases are important, as the differences in the shape behaviour of backscattered muons and protons, compared to the electrons and pions at different energies, arise from their different energy loss mechanisms within the matter - these differences give rise to the differentiation power of the technology. Electrons and pions have a lower threshold energy for production, allowing them to be produced and detected more efficiently at lower energies compared to muons and protons. Additionally, they lose energy more rapidly through processes such as bremsstrahlung and ionisation as they propagate through the Lunar regolith and get backscattered, leading to a more rapid decrease in their energy. Furthermore, distributional differences can be extracted from the time component (also showing a requirement of roughly 2 to 5 ns time resolution), as well as the scattering angle, providing a good measure of differentiation for particle types based on the interaction modes with the matter of varying density. 

We observe distinct patterns in the distributions of various particle properties. Notably, muons exhibit a unique behaviour, characterised by a flat distribution in both energy and time domains. This observation suggests a relatively consistent energy deposition within the detector's sensitive volume and a uniform temporal profile for these particles. The distinctiveness of muons' energy and time distributions might be attributed to their well-established penetration capabilities and their limited interactions with the detector material. In contrast, electrons exhibit markedly different distribution characteristics. These distributions manifest as narrow, well-defined energy, time, and scattering angle peaks. This trend is particularly pronounced when comparing electrons with heavier particles, such as muons, pions, and protons. The narrower peaks indicate that electrons deposit energy within a more confined range and exhibit tightly clustered arrival times, indicative of more predictable interactions and decays.
Moreover, the scattering angle distribution for muons underscores their distinct behaviour. The narrower scattering angle distribution suggests that muons are more likely to undergo minimal scattering compared to heavier particles. This behaviour aligns with expectations based on the interplay between particle mass, charge, and the scattering process. 

This study unveils intriguing variations between distributions obtained under different scenarios. Specifically, we observe discrepancies at both the peaks and tails of the distributions for these two scenarios. This divergence could be attributed to varying experimental conditions, material properties, or detector responses.

\begin{figure}[H]
\centering
\begin{subfigure}{0.33\textwidth} 
\includegraphics[height=5cm, width=\textwidth]{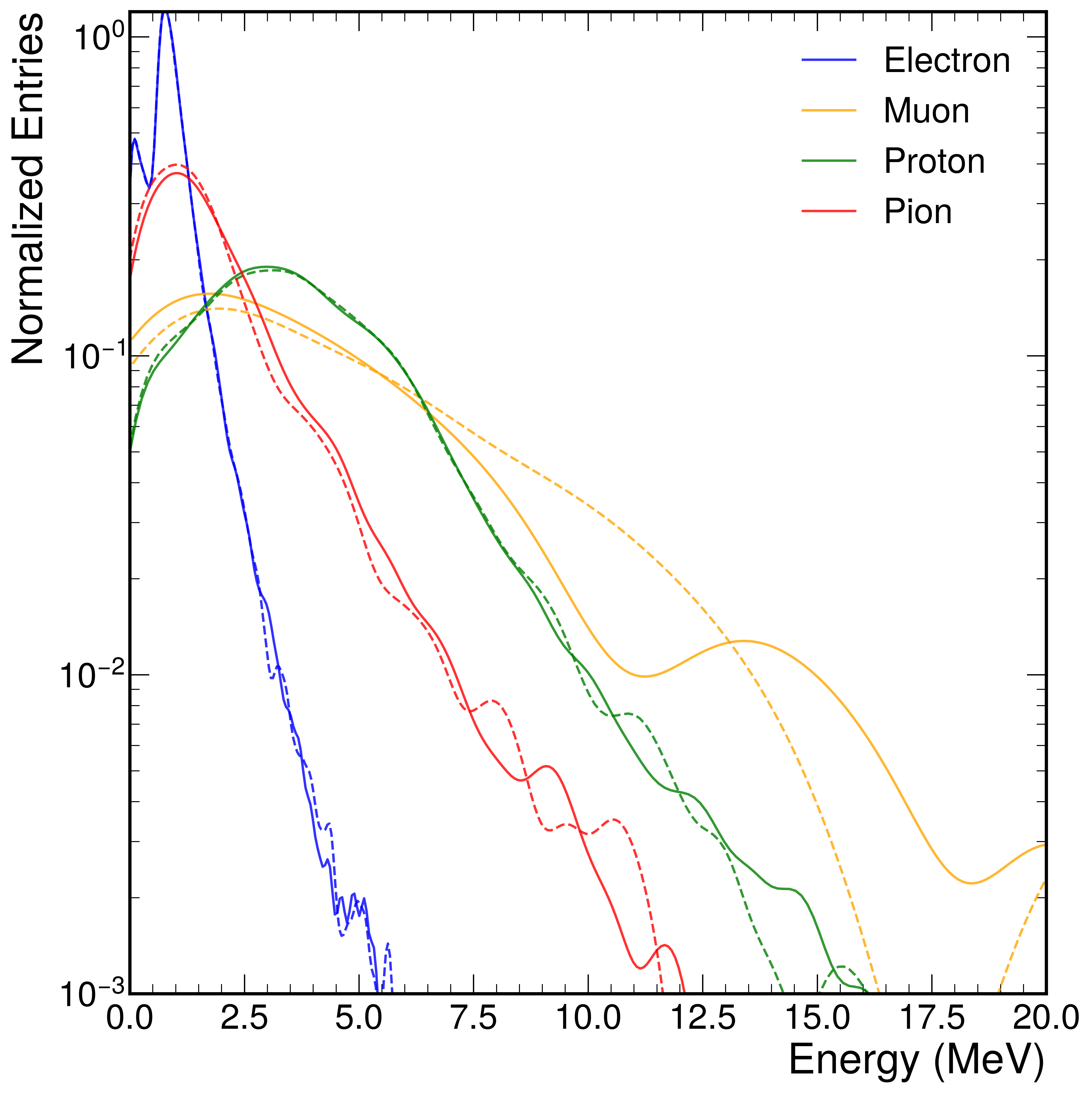}
\caption{Energy Deposited - Lunar}
\label{subfig:LunarEnergy_Spec}
\end{subfigure}%
\begin{subfigure}{0.33\textwidth}
\includegraphics[height=5cm, width=\textwidth]{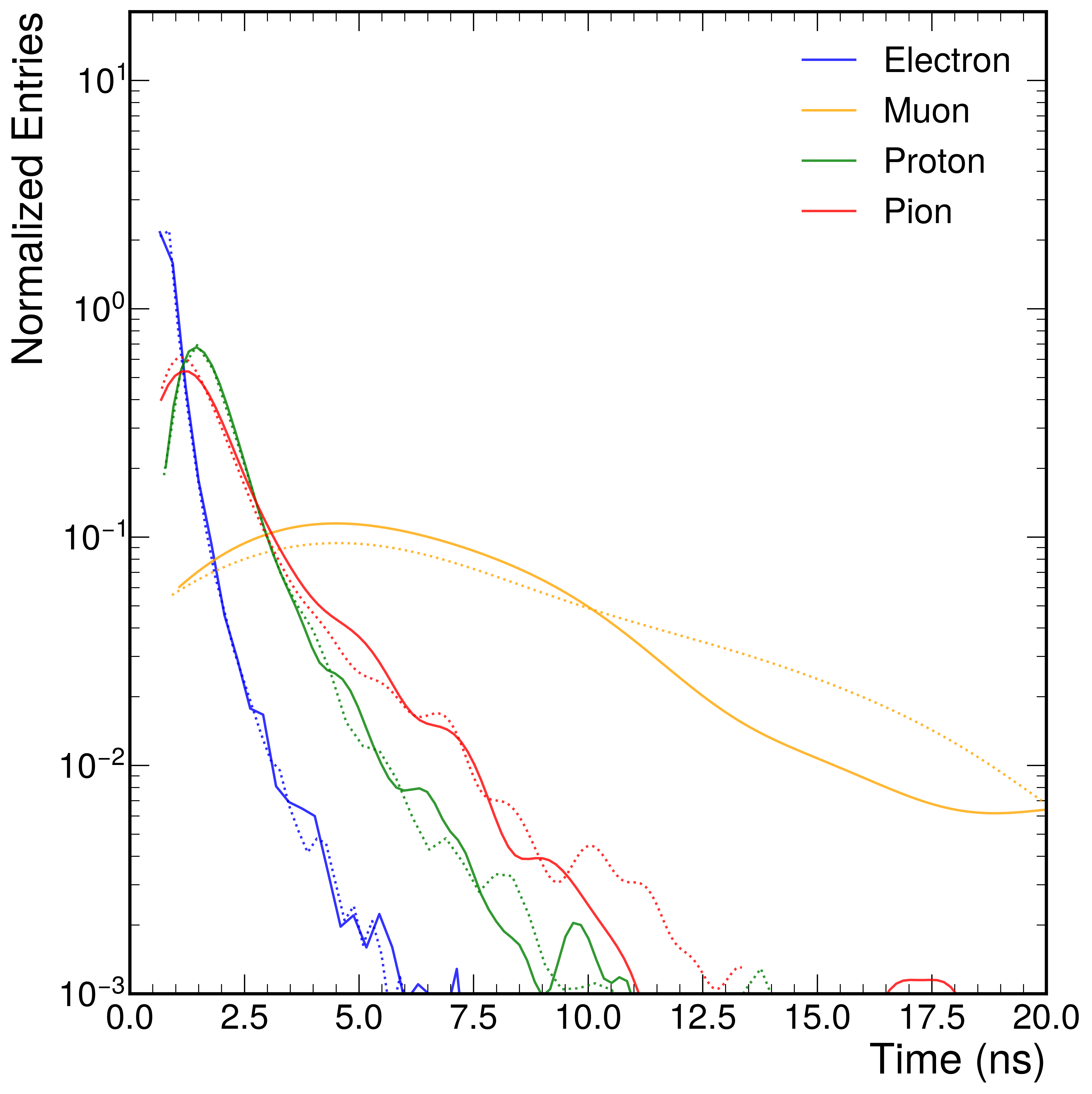}
\caption{Time Difference - Lunar}
\label{subfig:LunarTime_Spec}
\end{subfigure}%
\begin{subfigure}{0.34\textwidth}
\includegraphics[height=5cm, width=\textwidth]{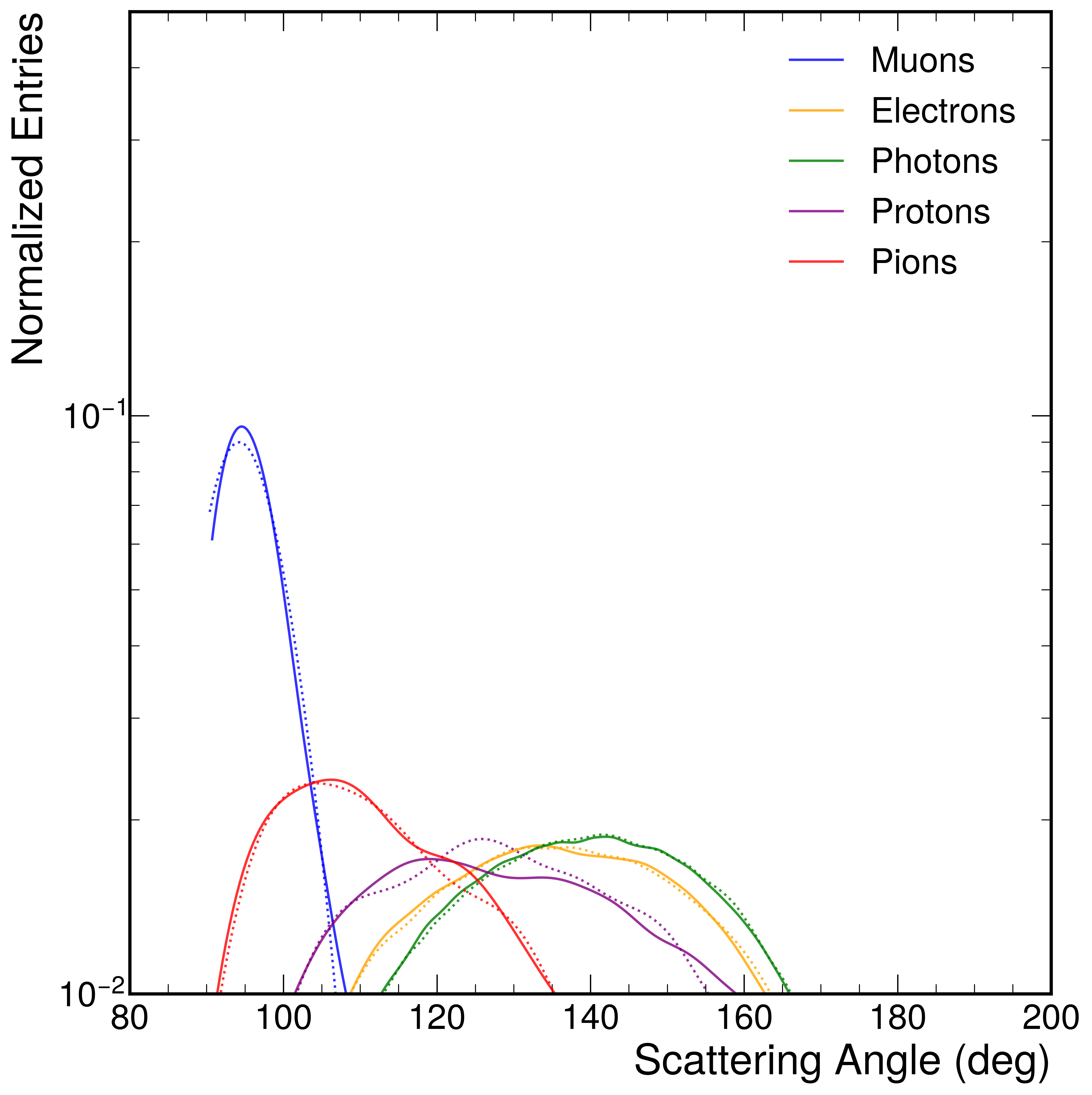}
\caption{Scattering Angle - Lunar}
\label{subfig:LunarAngle_Spec}
\end{subfigure}
\begin{subfigure}{0.33\textwidth} 
\includegraphics[height=5cm, width=\textwidth]{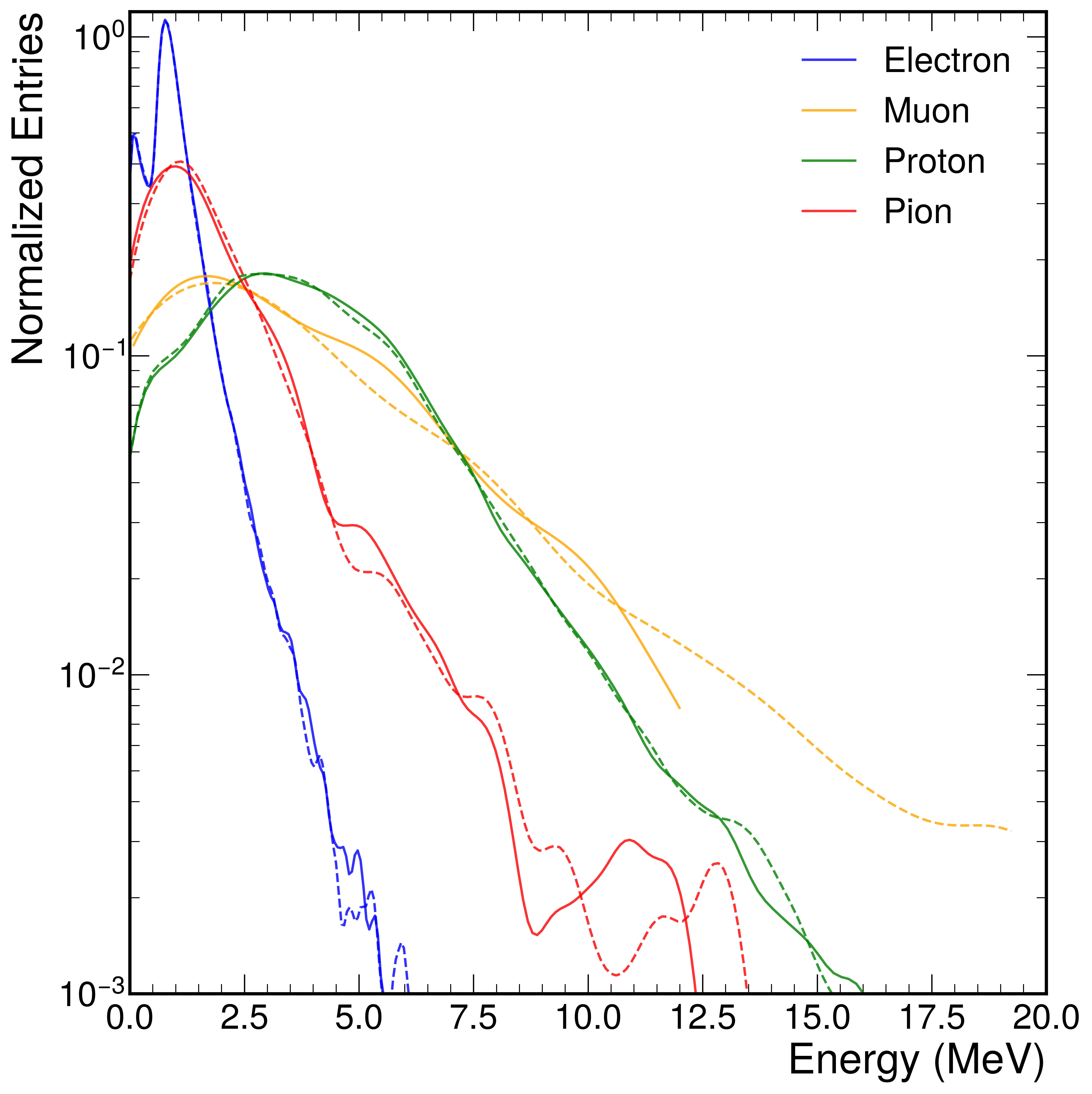}
\caption{Energy Deposited - Mars}
\label{subfig:MarsEnergy_Spec}
\end{subfigure}%
\begin{subfigure}{0.33\textwidth}
\includegraphics[height=5cm, width=\textwidth]{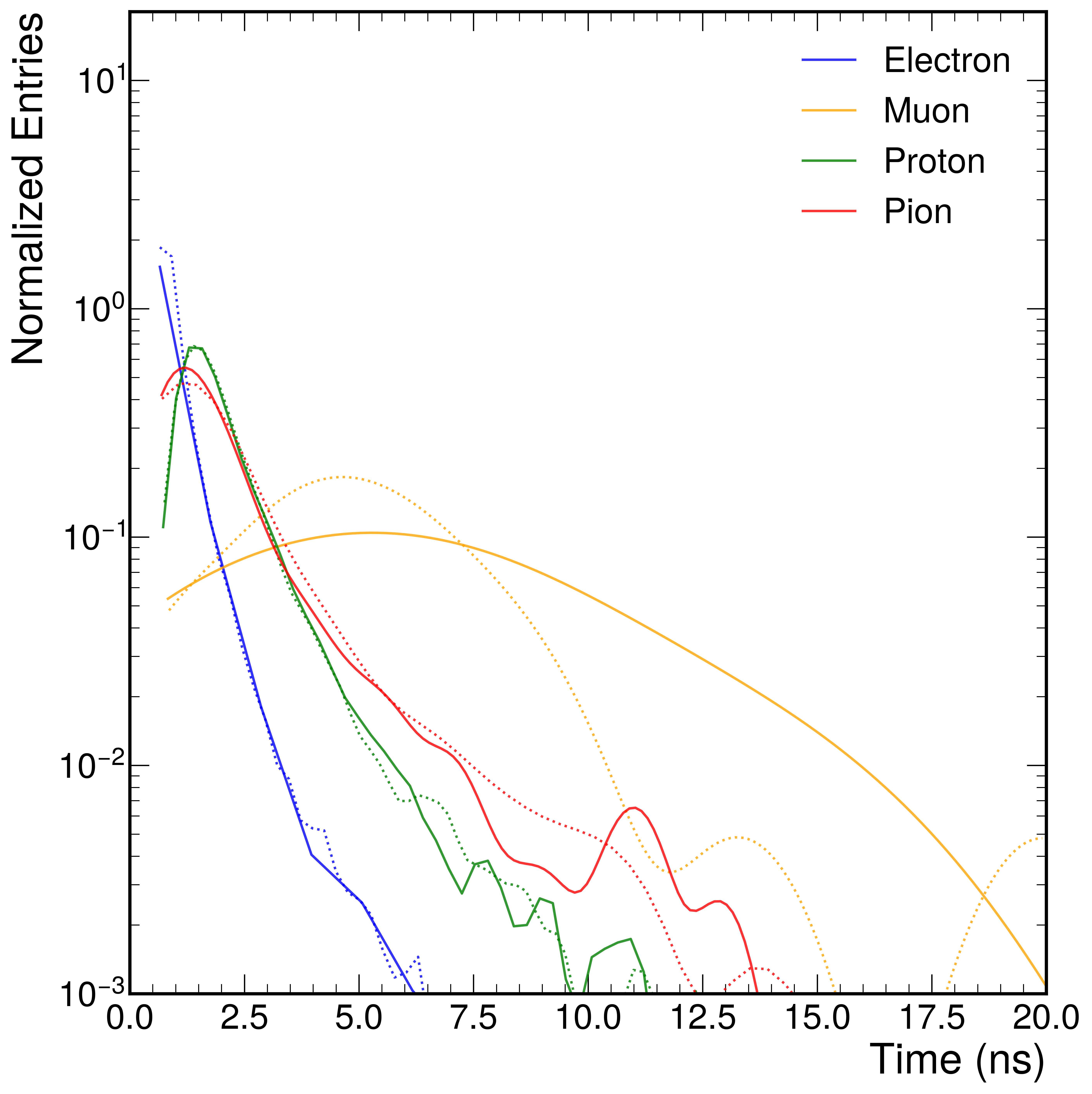}
\caption{Time Difference - Mars}
\label{subfig:MarsTime_Spec}
\end{subfigure}%
\begin{subfigure}{0.33\textwidth}
\includegraphics[height=5cm, width=\textwidth]{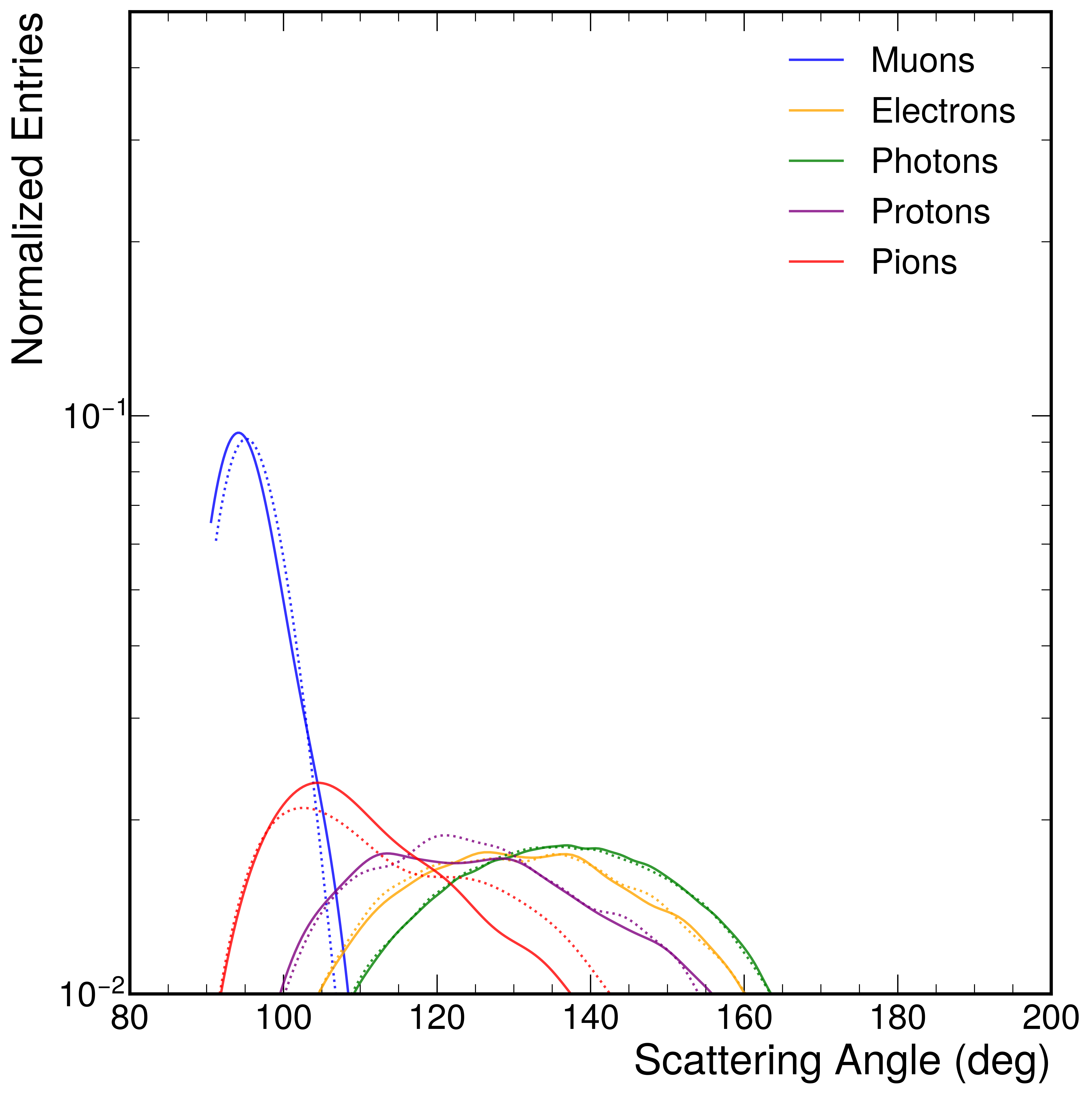}
\caption{Scattering Angle - Mars}
\label{subfig:MarsAngle_Spec}
\end{subfigure}

\caption{Comparison of distributions for both Lunar and Martian data, detailing the energy deposited (a,d), the time difference between detector planes (b,e), and the scattering angle (c,f). These distributions are obtained from various particles and are contrasted between Ice (represented by solid lines) and Rock (represented by dashed lines) datasets for each celestial body.}
\label{fig:EnergySpectrum}
\end{figure}

\begin{figure}[H]
\centering
\includegraphics[width=0.55\textwidth]{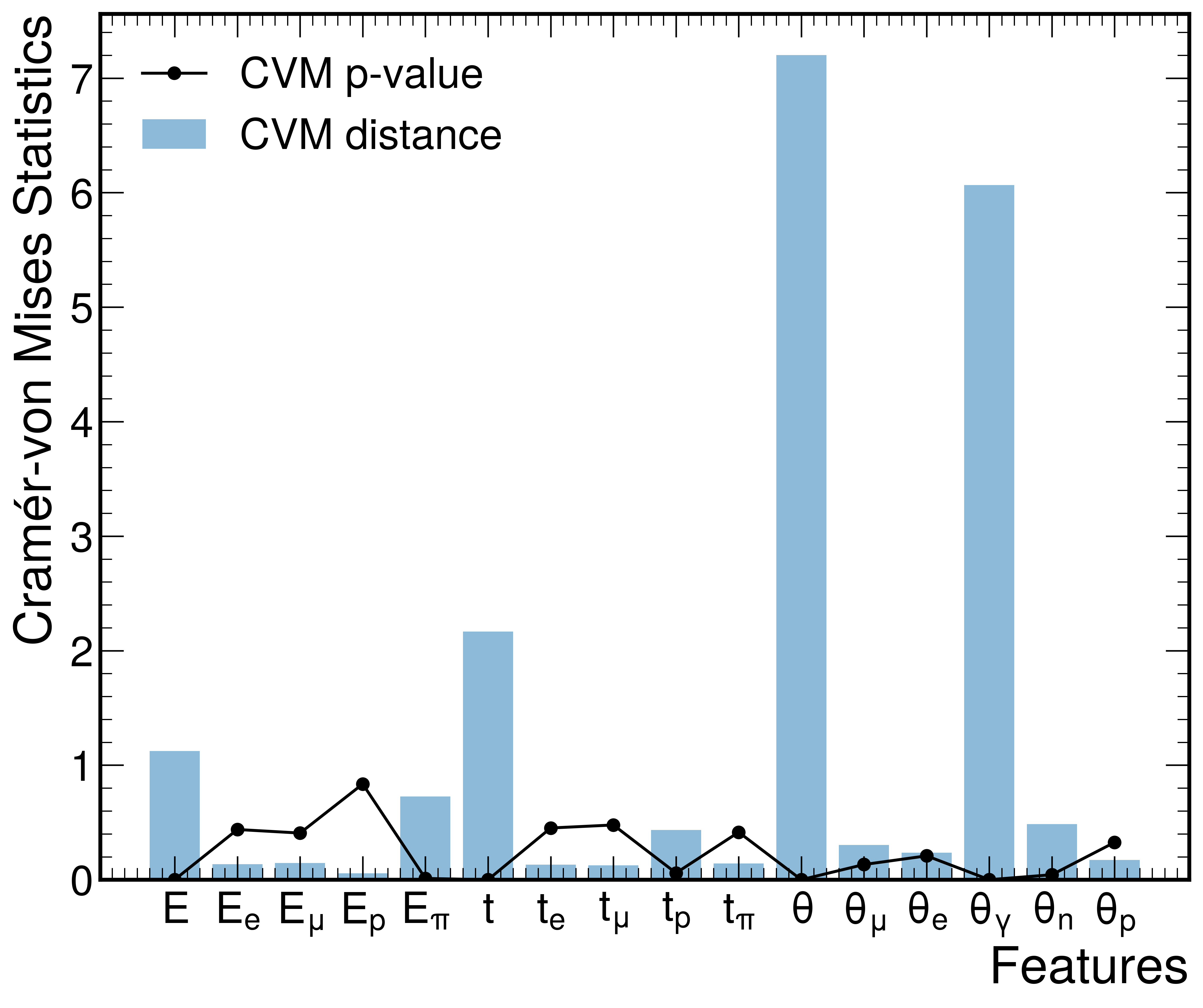}
\caption{Cramér-von Mises test results for various features in the case of Moon scenario.}
\label{fig:CVM_Distance}
\end{figure}

The Cramér-von Mises (CvM) test was employed to evaluate the similarities and differences between the distributions of specific parameters, such as energy, time, and angular measures, in rock and ice datasets, as depicted in figure~\ref{fig:CVM_Distance}. The CvM distance quantifies the divergence between the two distributions, while the p-value provides an indication of the strength of evidence supporting the hypothesis that the two samples originate from different distributions. The results show a substantial difference in the overall energy and a notable difference in pion energy, while other energy subcategories showed no significant differences. A highly significant discrepancy in the overall time was observed, with only a marginal difference in proton time. In terms of angular parameters, an extremely significant difference in the overall angular and backphotons was detected, with a significant difference in backprotons. Other subcategories of angular parameters and a few time subcategories did not exhibit significant differences. These findings offer detailed insights into the distinctiveness and similarities of the examined parameters, highlighting areas that may be pivotal in understanding the underlying differences between rock and ice datasets. 

\subsection{Machine Learning using Simulated Data}

In order to determine between the two simulated scenarios - ``Rock'' and ``Ice'', machine learning (ML) classification is applied. The use of ML in this context is rooted in its ability to handle complex patterns within large datasets efficiently. Specifically, an ML-based classifier, when optimally constructed, can process and label data with speed and accuracy, accommodating the intricacies of the underlying physical phenomena. For the given task, which is binary in nature, the decision tree (DT) family of methods is well-suited. Decision trees are non-parametric supervised learning methods that enable the creation of a hierarchical structure of decision nodes, performing binary tests based on the provided set of attributes. In this particular case, the random forest iteration of the DT family was applied~\cite{liu2012new}, a method composed of multiple decision trees working in concert, aptly suited for handling the multidimensional data input. In this classification, the data was split into 60\% training, 20\% validation, and 20\% test sets.
% , consisting of a total of \textbf{0.04\%} events/muons.

\smallbreak
 
\begin{figure}[H]
\centering
\begin{subfigure}{0.45\textwidth}
\includegraphics[height=6.cm, width=\textwidth]{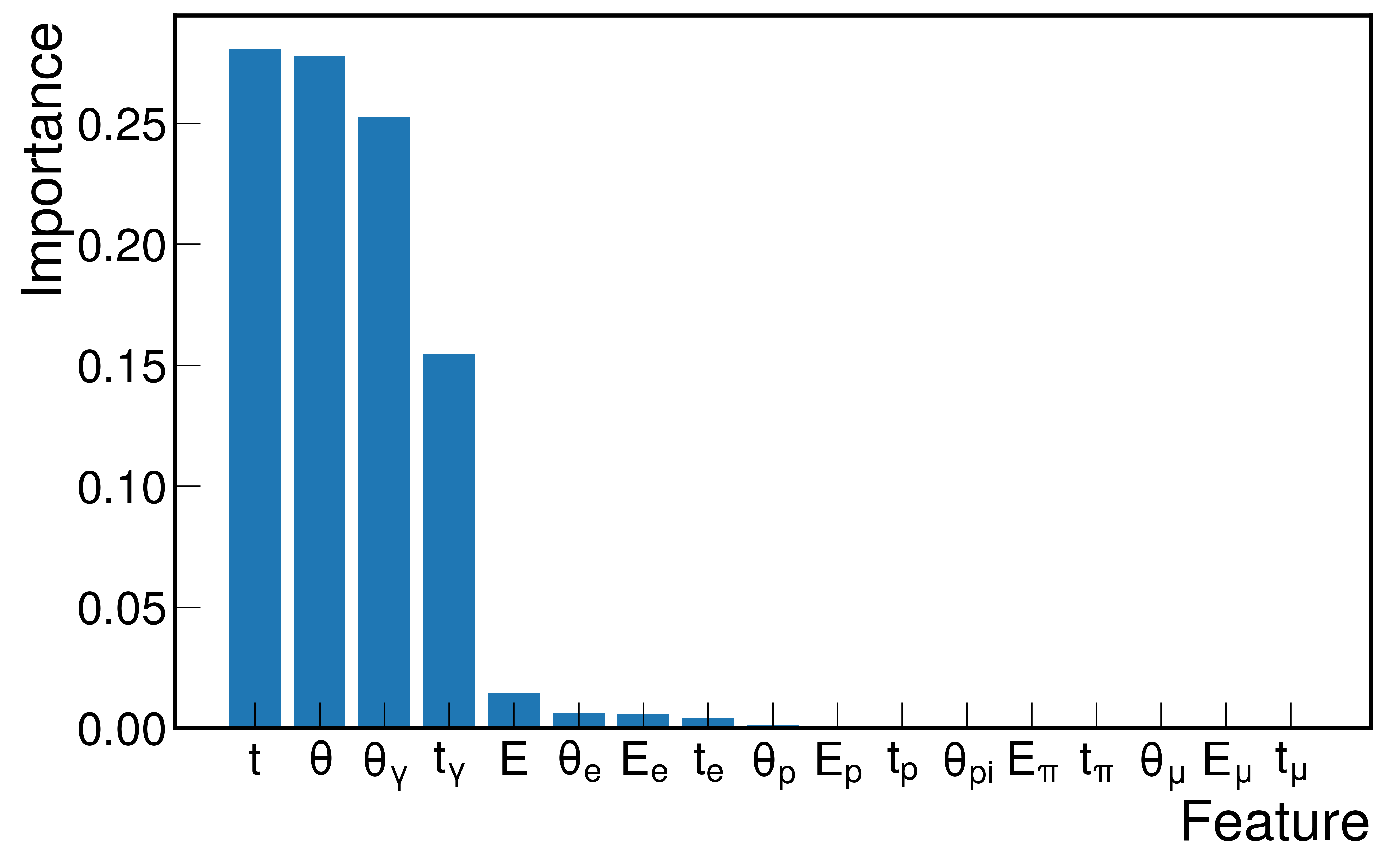}
\caption{}
\label{fig:Feauture_Imp}
\end{subfigure}
\begin{subfigure}{0.45\textwidth}
\includegraphics[height=6.cm, width=\textwidth]{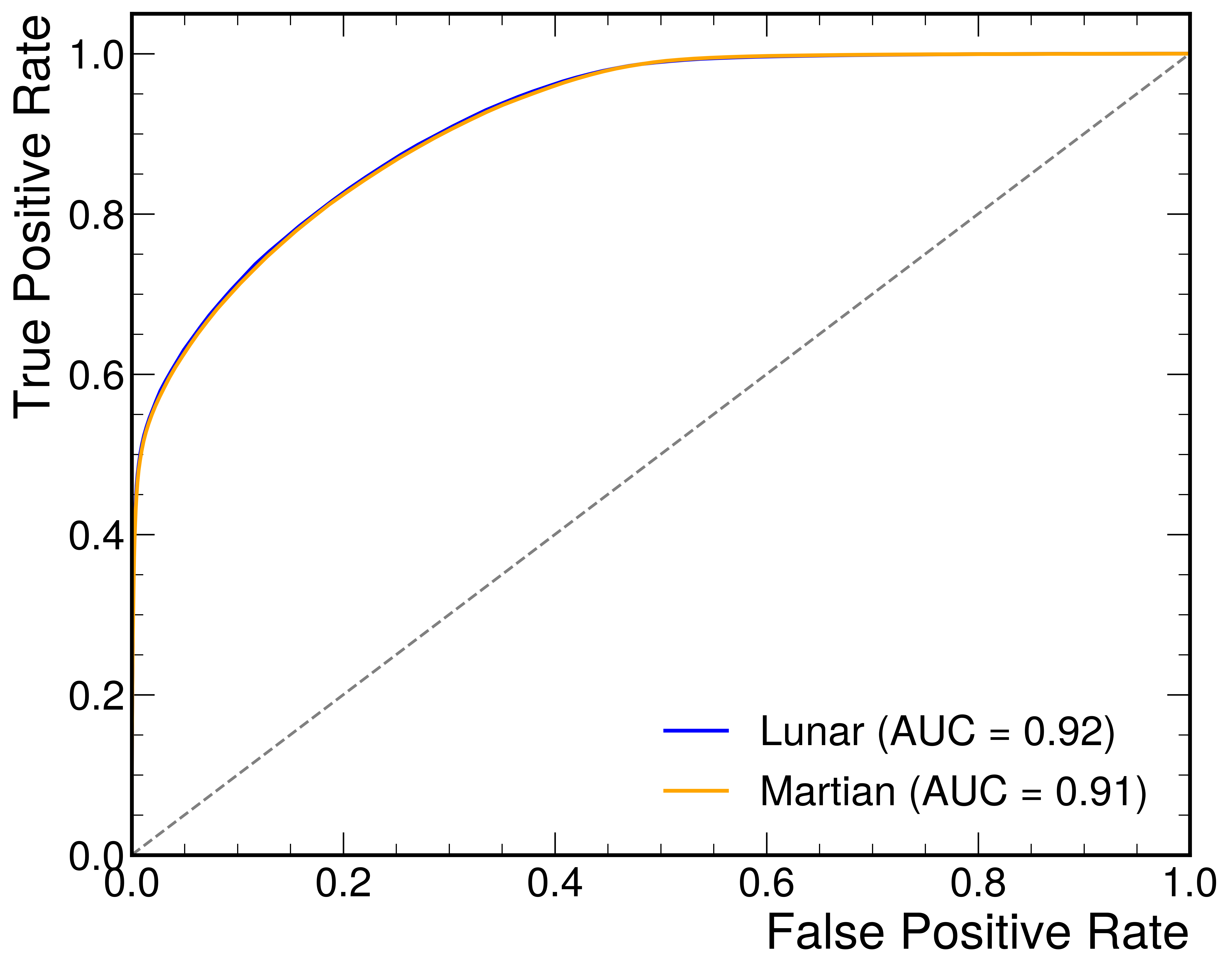}
\caption{}
\label{fig:ROC_Curve_MoonMars}
\end{subfigure}
\caption{a) Relative importance of the data features in the case of Moon. b) Comparison of Receiver Operating Characteristic (ROC) Curves for Ice vs Rock Classification on both Lunar and Martian Surfaces.}
\end{figure}

In the specific context of identifying water content in the subsurface, the ML model can analyse numerous features extracted from the simulation results. Through a process known as ``feature importance determination'', the ML model can identify the features carrying the most significant information for making accurate predictions in classifying between ice (signal) and rock (background). The feature importance for various parameters obtained is shown in figure~\ref{fig:Feauture_Imp}. Based on this analysis, the top 5 features, namely hit time, scattering angle (all particles), scattering angle contributed from photons, time contribution from photons, and energy (all particles), were chosen as they are the most influential in differentiating between ice and rock. These five features were utilised for training and classifying the data, laying the foundations for an efficient system that maximises the result with the least amount of computational power.

Having chosen the features to be included in the training and testing of the model, the classification performance of it has to be quantified in some way. One of the possibilities for it is calculating the Receiver Operating Characteristic Curve (ROC Curve) - a measure of the classifier's diagnostic ability at varying discrimination thresholds. By evaluating the True Positive Rate (positive data points classified as positive) and False Positive Rate (negative data points classified as positive) of the model, the Area Under the Curve (AUC) can be determined. AUC further quantifies the strength of distinguishing between the two classes that are being classified - the closer the AUC value to 1, the better the classification performance. In this study, both the Lunar and Martian scenarios were evaluated as shown in figure~\ref{fig:ROC_Curve_MoonMars}. For the Lunar case, the obtained AUC (Area Under the Curve) value of 0.92 (92\%) implies that the classifying performance of the developed model was very high, capable of efficiently differentiating between the simulated ice and rock scenarios. Similarly, for the Martian scenario, the model achieved an AUC value of 0.91 (91\%), demonstrating robust classification performance in distinguishing between Martian ice and rock formations.

\section{Conclusion \& Outlook}
This study highlighted the potential of muon tomography for space exploration, specifically for investigating water-ice content on the Lunar and Martian terrains. Through GEANT4 simulations, the study differentiates between dry and wet lunar surfaces through the analysis of backscattered particles. Moreover, the integration of machine learning offers an innovative approach to distinguish between geological formations such as ice from rock, emphasizing the technique's precision and reliability.

Initial simulations, however, highlighted challenges in capturing the Lunar/Martian topological and environmental nuances accurately. Future research aimed at these limitations can refine the findings. Enhancements, such as using scintillating fibres and more detector layers, aim to boost sensitivity and resolution. Transitioning from simulations to real-world experiments introduces complexities, including detector discrepancies, cosmic ray variations, and the intricacies of machine learning and geological interpretation. The application of digitization techniques, although slightly affecting the ROC curve's performance, augments the model's realism. Refinements in digitization and inclusion of parameters like track lengths hold promise for improving classification accuracy.

The datasets and analytical scripts used in this study have been made publicly available on Zenodo and can be accessed for further research~\cite{pinto2023detecting}.

\section*{Conflict of Interest}
The author declare that there are no conflicts of interest regarding the publication of this paper.

\section*{Acknowledgements}
This work is supported in part by the European Space Agency (ESA), under project code ESA CTR No. 4000139808/22/NL/MH/rp. The author is grateful for their valuable contribution.

\bibliographystyle{unsrt}

\begin{thebibliography}{99}
\bibitem{binder1998lunar} A. B. Binder, Science {\bf 281}, 1475--1476 (1998).
\bibitem{priyadarshini2009water} S. Priyadarshini, Water on Moon, Nature Publishing Group, (2009).
\bibitem{colaprete2010detection} Anthony Colaprete, Peter Schultz, Jennifer Heldmann, Diane Wooden, Mark Shirley, Kimberly Ennico, Brendan Hermalyn, William Marshall, Antonio Ricco, Richard C Elphic, and others, \textit{Detection of water in the LCROSS ejecta plume}, \textit{Science}, \textbf{330}(6003), 463-468 (2010), American Association for the Advancement of Science.

\bibitem{honniball2021molecular}
C. I. Honniball, P. G. Lucey, S. Li, S. Shenoy, T. M. Orlando, C. A. Hibbitts, D. M. Hurley, and W. M. Farrell, 
\emph{Molecular water detected on the sunlit Moon by SOFIA}, 
\emph{Nature Astronomy}, 
vol. 5, no. 2, pp. 121--127, 
2021, 
Nature Publishing Group UK London.

\bibitem{nazari2020water}
Mohammad Nazari-Sharabian, Mohammad Aghababaei, Moses Karakouzian, and Mehrdad Karami, 
\emph{Water on Mars—a literature review}, 
\emph{Galaxies}, 
vol. 8, no. 2, pp. 40, 
2020, 
MDPI.

\bibitem{gscan1}
G. Anbarjafari et al., 
\emph{Atmospheric ray tomography for low-z materials: implementing new methods on a proof-of-concept tomograph.}, 
\emph{arXiv:\textbf{2102.12542}}, 
2021.

\bibitem{pagano2021ecomug}
D. Pagano, G. Bonomi, A. Donzella, A. Zenoni, G. Zumerle, and N. Zurlo,
\emph{EcoMug: an Efficient COsmic MUon Generator for cosmic-ray muon applications},
\emph{Nuclear Instruments and Methods in Physics Research Section A: Accelerators, Spectrometers, Detectors and Associated Equipment},
vol. 1014, pp. 165732,
2021,
Elsevier.

\bibitem{ginzburg2013origin}
Vitali{\u\i} Lazarevich Ginzburg and Sergei Ivanovich Syrovatskii,
\emph{The origin of cosmic rays},
2013,
Elsevier.

\bibitem{anderson1979analysis}
DM Anderson and AR Tice,
\emph{The analysis of water in the Martian regolith},
\emph{Journal of Molecular Evolution},
vol. 14, pp. 33--38,
1979,
Springer.

\bibitem{ming2017chemical}
DW Ming and RV Morris,
\emph{Chemical, mineralogical, and physical properties of Martian dust and soil},
in \emph{Dust in the atmosphere of Mars and its impact on human exploration workshop},
number={JSC-CN-39581},
2017.

\bibitem{brun1997root}
Rene Brun and Fons Rademakers,
\emph{ROOT—An object oriented data analysis framework},
\emph{Nuclear Instruments and Methods in Physics Research Section A: Accelerators, Spectrometers, Detectors and Associated Equipment},
vol. 389, no. 1-2, pp. 81--86,
1997,
Elsevier.

\bibitem{taylor2016lunar}
Stuart Ross Taylor,
\emph{Lunar science: A post-Apollo view: Scientific results and insights from the lunar samples},
2016,
Elsevier.

\bibitem{agostinelli2003geant4}
Sea Agostinelli et al.,
\emph{GEANT4—a simulation toolkit},
\emph{Nuclear Instruments and Methods in Physics Research Section A: Accelerators, Spectrometers, Detectors and Associated Equipment},
vol. 506, no. 3, pp. 250--303,
2003,
Elsevier.

\bibitem{liu2012new}
Yanli Liu, Yourong Wang, and Jian Zhang,
\emph{New machine learning algorithm: Random forest},
in \emph{Information Computing and Applications: Third International Conference, ICICA 2012, Chengde, China, September 14-16, 2012. Proceedings 3},
pp. 246--252,
2012,
Springer.

\bibitem{pinto2023detecting}
Pinto, Olin Lyod,
\emph{Detecting Lunar and Martian Water via Backscattered Cosmic Particles using Muon Tomography},
August 2023,
Zenodo,
doi: 10.5281/zenodo.8343676,
url: \url{https://doi.org/10.5281/zenodo.8343676}.
\end{thebibliography}

\end{document}